\documentclass[aps,pra,twocolumn,superscriptaddress,preprintnumbers,longbibliography]{revtex4-2}

\usepackage[colorlinks=true, allcolors=blue]{hyperref}
\usepackage{graphicx}
\usepackage{dcolumn}
\usepackage{bm}
\usepackage{amsmath}
\usepackage{comment}

\begin{document}

\preprint{...}

\title{\textbf{Recommendations of the working-group on nuclear charge radii data} }
\title{\textbf{Towards improved determination and compilation of nuclear charge radii} }
\title{\textbf{Towards better nuclear charge radii} }

\author{István Angeli}\affiliation{Department of Experimental Physics, University of Debrecen, 4026, Hungary}

\author{Dimiter L. Balabanski}\affiliation{Extreme Light Infrastructure - Nuclear Physics (ELI-NP), Horia Hulubei National Institute for R\&D in Physics and Nuclear Engineering (IFIN-HH), 077125 Bucharest-Magurele, Romania}

\author{Paraskevi Dimitriou}\affiliation{International Atomic Energy Agency (IAEA),  Vienna International Centre, A-1400 Vienna, Austria}

\author{Dipti}\affiliation{Department of Physics and Astronomy, Clemson University, Clemson, SC 29631, USA}

\author{Kieran T. Flanagan}\affiliation{Department of Physics and Astronomy, University of Manchester, Manchester, M13 9PL, UK}

\author{Georgi Georgiev}\affiliation{Universit\'e Paris-Saclay, CNRS/IN2P3, IJCLab, 15 Rue Georges Clemenceau, 91400 Orsay, France}

\author{Mikhail Gorchtein}\affiliation{Institut f\"ur Kernphysik, Johannes Gutenberg-Universit\"at Mainz, 55128 Mainz, Germany}
\affiliation{PRISMA$^+$ Cluster of Excellence, Johannes Gutenberg-Universit\"at Mainz, 55128 Mainz, Germany}

\author{Paul Gu\`eye}\affiliation{Facility for Rare Isotope Beams, Michigan State University, 640 South Shaw Lane, East Lansing, MI 48824, USA}

\author{Fabian Heiße}\affiliation{Max-Planck-Institut für Kernphysik, Heidelberg, Germany}

\author{Andreas Knecht}\affiliation{Paul Scherrer Institute, Forschungsstrasse 111, 5232 Villigen PSI, Switzerland}

\author{Kei Minamisono}\affiliation{Facility for Rare Isotope Beams, Michigan State University, 640 South Shaw Lane, East Lansing, MI 48824, USA}

\author{Wilfried Nörtershäuser }\affiliation{Institut f\"ur Kernphysik, Technische Universit\"at Darmstadt, 64289 Darmstadt, Germany}
\affiliation{Helmholtz Research Academy Hesse for FAIR, GSI Darmstadt, Planckstr. 1, 64291 Darmstadt, Germany}

\author{Ben Ohayon}\email{Corresponding author: bohayon@technion.ac.il}\affiliation{The Helen Diller Quantum Center, Department of Physics,
Technion-Israel Institute of Technology, Haifa, 3200003, Israel}

\author{Natalia S.~Oreshkina}\affiliation{Max-Planck-Institut für Kernphysik, Heidelberg, Germany}

\author{B. K. Sahoo}\affiliation{Atomic, Molecular and Optical Physics Division, Physical Research Laboratory, Navrangpura, Ahmedabad 380009, Gujarat, India}

\author{Hunter Staiger}\affiliation{Department of Physics and Astronomy, Clemson University, Clemson, SC 29631, USA}

\author{Endre Takacs}\affiliation{Department of Physics and Astronomy, Clemson University, Clemson, SC 29631, USA}

\author{Xiaofei Yang}\affiliation{School of Physics and State Key Laboratory of Nuclear Physics and Technology, Peking University, Beijing 100871, China}

\author{Deyan T. Yordanov}\affiliation{Universit\'e Paris-Saclay, IJCLab, 15 Rue Georges Clemenceau, 91400 Orsay, France}\affiliation{Experimental Physics Department, CERN, 1211 Geneva 23, Switzerland}

\collaboration{Working Group on Nuclear Charge Radii}

\date{\today}

\begin{abstract}
Nuclear charge radii constitute a physical observable of growing significance across multiple subdisciplines of physics and related fields. Their determination relies on a combination of complementary experimental techniques and advanced theoretical frameworks.
Current recommended values are informed by the outcomes of several independent working groups, each employing distinct methodological approaches and evaluation strategies. 
The present effort is directed toward a more precise and reliable extraction of charge radii, as well as the development of a modern, transparent, and methodologically robust compilation of recommended values.
\end{abstract}

\maketitle


\section*{Preface}
In 2025, the IAEA organized a technical meeting on the “Compilation and Evaluation of Nuclear Charge Radii”, which focused on revising the table of recommended nuclear charge radii by Angeli and Marinova (2013). Since its publication, there have been significant developments in experimental and theoretical approaches associated with the measurement and evaluation of nuclear charge radii. This report, authored by the attendees of the meeting, presents a summary of the current status and key recommendations for future evaluations.

\section{Introduction}
The charge density within nuclei is relatively constant in the bulk volume and rapidly decreases beyond it. 
Thus, nuclei have a characteristic size that could be studied through electromagnetic probes, such as bound or scattered electrons and muons.
Electron-scattering experiments demonstrated that the root-mean-squared (rms) charge radii scale with the nuclear mass as $R_\mathrm{c}\equiv\sqrt{r_c^2} \propto  A^{1/3}$. 
From a nuclear structure point of view, deviations from this simple dependence are the features to be investigated since they reveal changes of nuclear properties. 

Detailed studies of changes in $R_\mathrm{c}$ and $\delta\left\langle r_\mathrm{c}^2\right\rangle$ along isotopic chains reveal the evolution of nuclear structure and provide information about collective properties and the underlying nuclear forces~\cite{Otten.1989}.
Their description is a stringent test of nuclear theoretical models~\cite{Reinhard.2017, 2023-Yang}. 

The $R_c$ observable is related to others that are sensitive to nuclear deformation, such as the nuclear electric quadrupole moment, $Q_0$, the reduced transition probability for an electric quadrupole transition between the ground state and the first excited $2_1^+$ state in even-even nuclei, $B(E2; 2^+ \rightarrow 0^+)$, see \textit{e.g.}~\cite{2024-Koszorus,2023-Yang,2025-Verney}.
Therefore, it is useful to consider the trends observed for nuclear moments when discussing evaluated values of charged radii since these quantities are strongly interrelated. 

Further insight into the fine details of the nuclear structure is offered by the mirror shift, \textit{i.e.} the charge radius difference between mirror partners $\Delta R_\mathrm{ch}\equiv R_c(N,Z)-R_c(Z,N)$. This is a useful observable that correlates with several interesting phenomena pertaining to nuclear structure. For example, it has been argued that $\Delta R_\mathrm{ch}$ 
which is correlated with the slope of the symmetry energy at the saturation density of nuclear matter~\cite{Brown.2017,Brown.2020,Konig.2024,Reinhard.2022}. Because $\Delta R_\mathrm{ch}$ is a small number resulting from subtracting two large uncorrelated ones, it has a large fractional uncertainty. Unlike isotope shifts, mirror shifts are particularly sensitive to the uncertainties in the individual radii.

Beyond their use in nuclear physics, charge radii are often required for tests of the Standard Model (SM). Among such precision tests are the superallowed nuclear beta decays that provide access to the top-left corner element of the Cabibbo-Kobayashi-Maskawa quark-mixing matrix $V_{ud}$~\cite{Hardy:2020qwl,Gorchtein:2023naa}, and the determination of the weak mixing angle with parity-violating electron scattering (PVES) off hydrogen~\cite{Qweak:2018tjf}. The aforementioned experiments are sensitive to heavy physics beyond SM (BSM) at scales ${\sim}{15}$ TeV and are  complementary to collider searches. In addition, low-energy experiments are naturally sensitive to hypothetical light BSM particles and long-range BSM interactions. 

The knowledge of the absolute and differential charge radii is the limitation for stringent tests of the SM and setting limits on hypothetical new physics via high-precision low energy experiments~\cite{2017-Delaunay,2025-gfactorRadii}. Additionally, the precise comparison of charge radii extracted by different methods offer new and unique theory insights~\cite{2022-Antognini,2024-RMPmuonic}.

%
PVES on heavy neutron-rich nuclei is used to extract the neutron skins of these nuclei~\cite{PREX:2021umo,CREX:2022kgg}, and precise knowledge of the respective charge radius is a prerequisite. The neutron skins of ${}^{208}$Pb and ${}^{48}$Ca allow one to draw conclusions about the properties of the neutron matter, shedding light on the QCD phase diagram, the early universe, and nuclear astrophysics. 

In the aforementioned precision tests with nuclei, the pertinent nuclear charge radii must be known to 0.1-0.01\%. 
The precision of the available theoretical tools, lattice QCD for the single nucleon case~\cite{FlavourLatticeAveragingGroupFLAG:2024oxs} and \textit{ab initio} nuclear theory for nuclei~\cite{Ekstrom:2022yea} is
currently limited to $\approx1\%$. Therefore, it is common to use more precisely known experimental nuclear charge radii as input.

In all the examples mentioned, the uncertainties of the nuclear charge radii, as well as other input parameters, are crucial. These uncertainties are frequently governed more by systematic, theory-driven effects than by experimental ones. In view of this, the reliability of the evaluation of nuclear radii plays a central role. To test our understanding of the theoretical errors and their reliability, it is important to confront and critically assess different methods of accessing nuclear radii. Historically, three methods have generally been used: spectroscopy of ordinary atoms, spectroscopy of muonic atoms, and electron scattering. As we will see below, the spectroscopy of highly charged ions is gaining momentum.

The last edition of the tables of nuclear charge radii by Angeli and Marinova dates back to 2013~\cite{Angeli:2013epw}, which is an update of Angeli's 2004 compilation~\cite{Angeli:2004kvy}. Other widely used tables are those by Fricke, Heilig, and collaborators from 1995 and 2004~\cite{Fricke:1995zz,Fricke:2004}. Another critical evaluation focuses on light nuclei~\cite{2025-Mirror}. These tables include all data from before their publication, as well as additional analysis done by the evaluators. The evaluators went to significant lengths to collect unpublished data through private communications and unpublished theses. When papers did not provide uncertainties for nuclear charge radii, or neglected major sources of uncertainty, uncertainties were estimated for the sake of the evaluation. Quantities such as the $V$ factors described in Section \ref{sec:combined_analysis}, nuclear polarization corrections, and field and mass shift coefficients were recalculated or interpolated when missing, to obtain the best nuclear radii possible.

To convert the absolute and relative radii measurements into a set of recommended nuclear radii,~\cite{Angeli:2004kvy} followed a weighted-least-squares optimization procedure that considered both isotonic and isotopic difference measurements. The procedure included consideration of external and internal uncertainties, as well as correcting theoretical field-shift coefficients using radius measurements from electron scattering and muonic atom spectroscopy. In contrast,~\cite{Fricke:2004} employed an element-by-element procedure using the Barrett moment recipe and a combined King Plot analysis. In general, both analyses showed good agreement with each other, but each contains details or small corrections that are difficult to understand from the published material. Clearly, in light of the high demand and much activity in the field, a comprehensive, up-to-date, transparent critical evaluation and compilation of nuclear charge radii is sorely needed.

This report is dedicated to providing a short overview of the current status, open problems, and recommendations from the community for future compilation. 
The first section deals with the workhorses of absolute radii determinations; elastic electron scattering, muonic atom spectroscopy, and their combination.
The second section deals with the widely-used method of obtaining the isotopic difference in charge radii through laser spectroscopy of natural atoms and singly-charged ions, as well as the atomic theory needed to apply it precisely.
The third section deals with the growing field of absolute and differential radii measurement using observables in multiply-charged ions.

\begin{figure*}[tb]
	\centering

\includegraphics[width=\linewidth,clip]{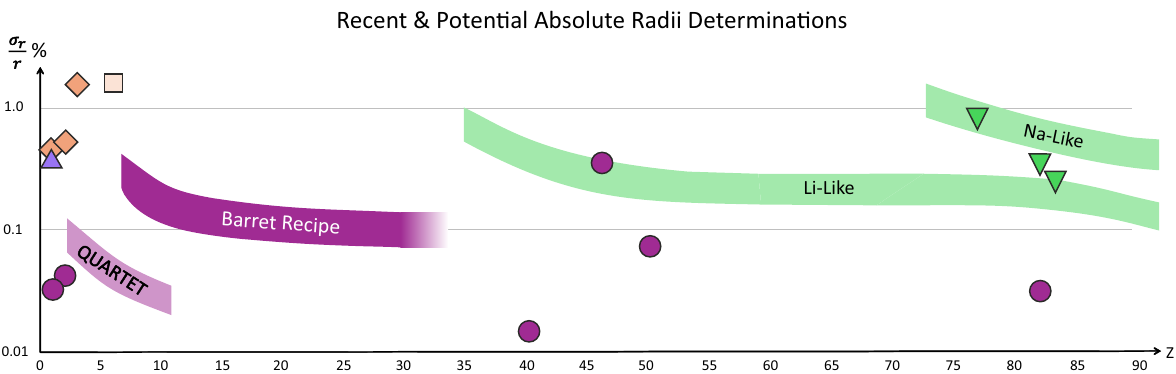}
\caption{Fractional uncertainty obtained in recent absolute radius determinations that include:
Hydrogen spectroscopy~\cite{2025-CODATA} (up-pointing triangle),
electron scattering~\cite{PhysRevC.77.041302, 2011-SickLi, 2019-protonScat} (diamonds), 
muonic atom energies~\cite{2024-RMPmuonic, 2025-Mirror, 2025_Sun, 2025-Zr, 2022-PdFermi} (circles),
and the spectroscopy of Na-like lithium-like ions~\cite{2025_Yerokhin,Staiger2025} (down-pointing triangles).
Some potential improved determinations based on the current status of the theory are indicated by light-shaded bands~\cite{2024-QUARTET, 2025-LiTh, PhysRevLett.134.123001, Gillaspy2013}.}
	\label{fig:Abs}
\end{figure*}

\section{Electron Scattering, muonic atoms, and their combined analysis}
\subsection{Electron Scattering~\label{sec:scattering}}
Historically, the first evidence for a finite spatial extension of the nuclear charge distribution was obtained through elastic electron scattering~\cite{Hofstadter:1956qs}. To this day, electron scattering remains an important source of information about the nuclear charge distribution~\cite{1987-VJV}.

Since the radius is defined as the slope of the form factor at zero momentum transfer, from the scattering perspective, this constitutes the principal difficulty in extracting the radius from the scattering data. At very low momentum transfer values, the numerical effect of the radius on the scattering cross section becomes prohibitively small, whereas at higher momentum transfer, it becomes intertwined with higher moments of the charge distribution and higher-order corrections. Of the latter, the two-photon exchange (TPE) in $ep$-scattering has attracted the most attention in the past two decades. This activity should be transferred to electron-nucleus scattering that involves TPE and virtual nuclear excitations (historically coined dispersion corrections).  Parameterization of the $Q^2$-dependence of the form factor can also be used to extract the rms~\cite{deVries87}. We refer the reader to the recent reviews on electron-proton~\cite{Afanasev:2023gev} and electron-nucleus~\cite{Gueye20} scattering and the references therein.

\subsection{Muonic atoms}
Both the experimental methods and the theoretical formalisms used for the calculations strongly depend on the nuclear charge number.
The radii of the lightest nuclei ($Z = 1$ and $Z=2$) have been extracted from laser spectroscopy of the so called `Lamb shift' ($2s-2p$ transitions). 
The most reliable values are given in~\cite{2024-RMPmuonic}. 
Reference radii of the $Z = 3-5$ elements cannot be accurately extracted from existing measured muonic atom energies due to their low precision; therefore, they are currently taken from scattering experiments of varying reliability (see the discussion in~\cite{2024-QUARTET}). Of these, only the scattering results in $^6$Li have been scrutinized~\cite{2011-SickLi}.
For $Z = 6-10$, the radii extracted from muonic atoms have comparatively large experimental errors (see e.g. Fig. 4 in~\cite{2025-Mirror}), making the tabulated radii probably quite reliable.
From $Z=11$ on, the accuracy of charge radii extracted from $np-1s$ muonic atom energy levels is primarily limited by the extraction procedure and not by the experiment.
However, the experimental resolution is typically too low to distinguish the fine structure between transitions involving, e.g., $2p_{1/2}$ and $2p_{3/2}$ states. 

Starting from middle-heavy nuclei, e.g., Zr ($Z=40$), theoretical predictions are becoming highly sensitive to the nuclear charge density distribution utilized. The modern standard is the use of the two-parameter Fermi (2pF) distribution. The separation between fine-structure lines is becoming larger than the experimental resolutions; however, the use of 2pF requires, strictly speaking, fitting two parameters—therefore, two experimental lines are not enough for two theory parameters. 
Therefore, the precise determination of nuclear rms requires more detailed experimental spectra, including transitions to $2s$ and $2p$ states, which are unfortunately typically excluded from any compilations.

\subsection{Nuclear shape and Combined Analysis\label{sec:combined_analysis}}
It is generally believed that interpreting electron scattering experiments in terms of absolute charge radii is rarely applicable when the accuracy goals fall below a few percent. This view is partially based on comparisons with extractions from muonic atoms and, recently, from the spectroscopy of highly charged ions~\cite{2025-Endre}.
Deviations by $1-2\%$ percent between the radii extracted from muonic atoms and those from scattering are shown in Table~VI of~\cite{1992-Fricke} and in recent work~\cite{2025-Cl}, already for relatively light nuclei. These deviations exceed the combined uncertainty and are thus ascribed to the missing systematic corrections in analyzing the scattering experiments (see section~\ref{sec:scattering}). 
%

Although radii from muonic atom energies are more accurate than those from scattering, they are constrained by our understanding of the nuclear model as well as dynamic effects, which worsen for heavy nuclei.
In light of this, the community has adopted a method to combine the two complementary approaches. 
Many model-independent charge distributions from the scattering side depend on measured muonic atom energies as constraints (See remark ``g" in Table~I of~\cite{1987-VJV}).
Energies from the muonic side are often analyzed using Barret-moments~\cite{1970-Barrett}, which are converted to charge radii via ratios of moments from electron scattering~\cite{1974-Engfer, Fricke:1995zz}.
These moments are designed to have reduced model-dependence; however, the extent remains unclear and requires investigation. 

This Barrett moment recipe produces radii more accurately than using only $np-1s$ muonic energies, though it has limitations.
Estimating the uncertainties and correlations of the ratios-of-moments ($V$-factors) is unclear. Although estimates in~\cite{2025-Mirror} are useful, they do not quantify uncertainty directly. Systematic errors in the measured ratios, such as those from neglecting deformation effects, are absent. 
Nevertheless, when comparing the results with the approach using multiple transitions, the resulting radii agree to better than $0.1\%$.

Beyond realistic uncertainties in $V$-factors, it is crucial to know how to proceed when these are unavailable, which occurs in the majority of instances. In Fricke's book~\cite{Fricke:2004}, the approach sometimes involved taking values from neighboring elements (e.g., Eu), using the same factor for all isotopes (e.g., Ru, Ba, and Sr), interpolating them linearly between isotopes (e.g., Ca), or assuming a Fermi distribution with a specific skin thickness (e.g., Xe).
These approximations are not currently included in the uncertainty budget for either the absolute or differential radii and are significant. 

A promising approach moving forward is to calculate $V$-factors using nuclear theory (see e.g.~\cite{2025-Cl}). Such calculations should be tested against reliable factors (with uncertainty in Table I of~\cite{2025-Mirror}). 

To summarize the current situation broadly (see Fig.~\ref{fig:Abs}), the accuracy of charge radii is poorer than approximately $1\%$, $0.3\%$, and $0.1\%$ based on electron scattering, single-transitions in muonic atoms, or combined analysis, respectively. Claiming any uncertainty significantly below these values requires a more advanced analytical procedure, as demonstrated in light systems~\cite{2024-RMPmuonic} and certain magic nuclei~\cite{2025_Sun, 2025-Zr}. 

\subsection{Recommendations}
\begin{itemize}

\item Do not average different radii extracted from various electron scattering experiments unless it is clear that the results are dominated by statistics, which is rarely the case.

\item Study, both with experiment and theory, the higher order corrections (e.g., QED and multi-photon exchange) to radii extractions from electron scattering~\cite{Gueye20}.

 \item Revisit the analysis and compilation of electron scattering data in deformed nuclei.

\item Measure and analyze the extraction of radii in electron scattering using both the $Q^2$-dependence and the slope of the form factors at low $Q^2$.

\item Calculate and measure the contribution from higher-order QED and dispersive corrections to scattering results. Regularly update the database accordingly.

\item Compare the scattering of unpolarized and polarized electron and positron beams off nuclei.

\item  Create an updated critical re-evaluation for the scattering data, with an estimation of QED and all neglected effects.

\item Update the radii of $Z=1$ and $Z=2$ nuclei extracted from laser spectroscopy of muonic atoms as QED corrections evolve. 

\item Measure, fit, and analyze additional muonic spectral lines beyond the $np-1s$ manifold.

\item Re-measure known muonic lines for which the measurements have not followed very rigorous standards, especially in important cases such as magic nuclei.

\item Include a covariance matrix to account for isotopic and other correlations between the nuclear polarization errors.

\item Report radii from muonic atoms and from scattering separately, without averaging them (as done in Ref.~\cite{Fricke:2004})

\item Investigate the error budget of the Barrett-moment recipe, e.g., by extending the ``all-muonic'' approach that fits several transitions to more cases and systematically comparing the two approaches.

\item Include the uncertainty in ratios of moments (``V-factors") in future compilations, along with its covariance matrix for isotopes.

\item Investigate the potential to calculate V-factors \textit{ab initio} by comparing them with well measured cases.

\item Include exact descriptions of calibration procedures in the publications. 

\item Make all relevant data available open access to allow for future re-analysis as procedures develop and theoretical calculations improve.

\end{itemize}

\section{Laser spectroscopy of atoms and ions}

\begin{figure*}[htb]
	\centering
	\includegraphics[width=0.99\linewidth]{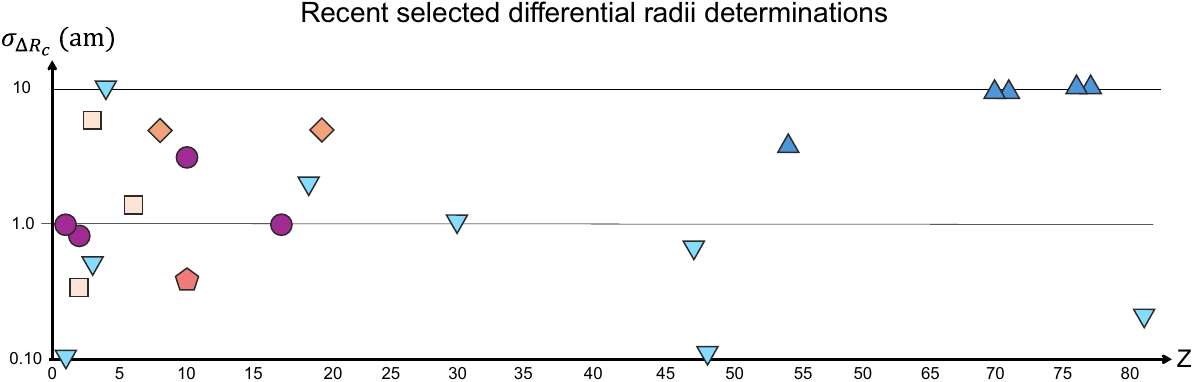}
	\caption{Uncertainty obtained in recent differential radius determinations between stable nuclei. 
    Some examples from dedicated electron scattering experiments~\cite{1979-O,1983-Ca} (diamonds),
    Isotope shifts in muonic atoms~\cite{Pohl.2010,2016-Pohl,Krauth.2021,2025-3He,2025-HeNuc, 1992-Fricke,2025-Mirror,2025-Cl, 2026-Hybrid, 2024-RMPmuonic} (circles),
    hydrogen and single-valence systems (atoms and singly-charged ions) with calculated atomic factors~\cite{2010-HD, 2018-3p, 2013-Li, 2012-12Be, Bijaya13, Bijaya16, Bijaya7,Zn-radii-3, Bijaya2, 2024-Zn, 2023-Cd, 2021-Bi} (down-pointing triangles), 
    Helium and helium-like ions~\cite{1994-Liplus, 2025-HeQun,2025-IonHE, PhysRevResearch.7.L022020,2025-HeMet, 2024-HeHyp, Muller.2025} (squares),
    the neon bound electron differential g-factor measurement~\cite{2022_Sailer} (octagon),
    Na- and Mg-like highly charged ions~\cite{Silwal2020, 2025-Endre, Hosier2025} for which the isotope shift is denoted by an upright triangle and the nuclide shifts by pairs of triangles that span both elements.}
	\label{fig:diff}
\end{figure*}

\subsection{Charge radius extraction}
Nearly all known radii of unstable nuclei have been obtained using isotope shifts (IS) determined with laser spectroscopy~\cite{Fricke:2004,Angeli:2013epw}.
The extraction of the change in the ms charge radius $\delta\left\langle r_\mathrm{c}^2\right\rangle^{A,A'}$ is based on the isotope shift in an atomic transition 
\begin{equation}\label{eq:IS}
    \delta\nu^{A,A'} = \nu^A-\nu^{A'}\approx K\,\frac{M_{A} - M_{A'}}{M_{A}M_{A'}} + F\, \Lambda^{A,A'}
\end{equation}
and it requires knowledge of the atomic mass-shift and field-shift factors $K$ and $F$, respectively. $M_A$ are the nuclear masses of the isotopes, and $\Lambda^{A,A'}$ is the change in the nuclear-size parameter in the lowest order $\Lambda^{A,A'}=\delta\langle r^2\rangle^{A,A'}$. 

The uncertainty contribution caused by the atomic factors often exceeds the statistical uncertainty from the experiment; see, for example,~\cite{Sc-radii-1, Sc-radii-2}, particularly far away from the stable isotopes and in regions with large variations of $\Lambda^{A,A'}$. Total charge radii $R_{A}= \sqrt{\langle r_\mathrm{c}^{2} \rangle_{A}} = \sqrt{R^2_{A_\mathrm{ref}}+\delta \left\langle r_\mathrm{c}^{2}\right\rangle^{A,A_\mathrm{ref}}}$ are further affected by the uncertainty of the reference radius, which is usually determined from muonic spectroscopy and/or electron scattering. 

Historically, there have been several sources of atomic factors. The extraction of the field-shift factor from the hyperfine structure or using the Goudsmit-Fermi-Segrè (GFS) rule~\cite{1958-Kopf} is no longer considered sufficiently accurate for high‑precision work and can provide, at best, an approximate or auxiliary method. Instead, \textit{ab initio} atomic structure calculations have reached at least comparable, often superior accuracy, as will be discussed in Sec.\,\ref{sec:atomic_factors}. For elements with at least three isotopes for which accurate charge radii are known from muonic atoms, electron scattering, or -- preferably -- the combined analysis, a King plot can be performed to determine the atomic factors.   

\subsection{Atomic Factors from a King-Plot Analysis}
A King-plot analysis~\cite{king} provides $K$ and $F$ factors for elements with at least three stable isotopes that have well known charge radii, e.g., $R_{\mu\mathrm{e}}$ from the combined analysis of electron scattering and muonic atom data, as, for example, in the case of nickel~\cite{Ni-radii-1, Ni-radii-2}. This approach can be hampered if the available charge radii from electron scattering and muonic atom spectroscopy have large uncertainties, as is the case, e.g., for the Zn isotopes~\cite{Zn-radii-1,Zn-radii-2,Zn-radii-3}. In some cases, there are outliers in the King plots that must be investigated closely. Reasons can include underestimated uncertainties of the charge radii or isotope shifts, as well as atomic-structure effects such as fine structure mixing, especially when it is enhanced by the presence of hyperfine structure. Odd isotopes of even-$Z$ elements are particularly prone to this effect. An example is the stable isotope $^{87}$Sr, which deviated from the line of even-even isotopes
unless second-order hyperfine structure effects are accounted for~\cite{Berengut2025}. If such an odd-isotope is relatively isolated from the other isotopes in the King plot, their position might have a strong influence or even practically determine the slope in the King plot and, therefore, the extracted $F$ and $K$.

Since the King plot is expected to be linear, with only very small deviations from higher-order effects, it can be used to obtain improved charge radii, even for the isotopes that have been used to determine the atomic parameters. This is regularly done, and the improved radii are traditionally denoted as $R_{\mu\mathrm{eo}}$ as they arise from the combined analysis from muonic, electronic, and optical data. We note that these radii should never be used as input values for another King plot, at least not without extreme care. 

Finally, we note that there are instances, especially in mono-isotopic elements with open atomic shells, where reliable atomic factors do not currently exist. Various approximations have been made to address this gap, such as estimating the NMS from the scaling law with SMS~$=0$ (e.g. in Cs and Tb~\cite{Fricke:2004}). Or calibrating on neighboring chains (e.g. in Nb~\cite{Fricke:2004}). The current compilations do not reflect the uncertainties associated with these approximations, and we suspect that they dominate the final uncertainty.  

\subsection{Calculated Atomic Factors}  
\label{sec:atomic_factors}
In most cases, a King plot extraction of both isotope shift factors is not possible or lacks accuracy. Thus, \textit{ab initio} calculations of these factors play an important role.
In particular, atomic calculations of these parameters become increasingly challenging when IS measurements are limited to a few isotopes~\cite{Bijaya1} or lack comparable precision among different transitions~\cite{Bijaya2,Bijaya3}. 
Conversely, comparing the calculated IS parameters with those inferred from the King plot provides a valuable test of the many-body methods employed~\cite{Bijaya5}.   

Field-Shift factors can often be obtained with reasonable accuracy using many-body methods~\cite{Bijaya1,Bijaya5,Bijaya6}, primarily because their operators have a  one-body form and their evaluation largely depends on the precision of the wave function in the nuclear region. Recent analyses further suggest that the calculation of these factors is relatively insensitive to the choice of the nuclear charge distribution model~\cite{Bijaya5,Bijaya7}.

A variety of relativistic many-body techniques—such as many-body perturbation theory, configuration interaction, random phase approximation, and coupled-cluster (CC) theory—have been employed in the literature to determine the isotope shift factors of atomic systems. Among these, the relativistic CC (RCC) method stands out as a particularly powerful approach due to its ability to rigorously incorporate both relativistic and electron correlation effects. However, current applications of RCC methods for calculating isotope shift factors are primarily limited to finite-field and analytical-response approaches~\cite{Bijaya5}, which present several challenges and limitations. These issues can be mitigated by utilizing alternative variants of RCC theory, such as unitary, equation-of-motion, and normal RCC methods, which offer promising avenues for more accurate and comprehensive calculations.


\subsection{Isomer Shifts}
Nuclear excitations also cause changes in the spatial distribution of protons that shift the energy levels in the atom. These ``isomer shifts''  are valuable to nuclear physics~\cite{2024-isomers}. Isomer shifts are largely free of the mass effect; therefore, following Eq.~\eqref{eq:IS}, they are proportional to the change in the isomeric charge-radius in the nucleus:
\[\delta\nu\approx F\delta\langle r_\mathrm{c}^2\rangle.\]
In past compilations, they were either simply acknowledged~\cite{Fricke:2004} or compiled relative to the most precise measurement in each isotopic chain~\cite{Angeli:2013epw}. 
When referencing isomer shifts to the nuclear ground state, the usually dominant uncertainty from the mass shift is canceled.

\subsection{Recommendations}

We conclude our discussion of the optical isotope shift with the following recommendations. 
To allow for future reanalysis when more accurate IS factors, reference charge radii, or theoretical results become available, it is essential that all underlying data be made accessible. Therefore, the following details should be explicitly presented and tabulated either in the main article, its supplementary information, or a persistent, publicly accessible database:
\begin{itemize}
    \item The measured isotope shifts with statistical and systematic uncertainties, information about possible correlations. 
    \item Source, values and uncertainties of reference radii and, if used in their derivation, $V$-factors.    
    \item Consider uncertainties in $V$-factors and their correlations when calibrating atomic factors using muonic data.
    \item Procedure used to perform the King plot, e.g., orthogonal distance regression, York fit, Monte Carlo method, etc.   
    \item Results of the King-plot, $K$ and $F$ factor of all included transitions including the correlation matrix.
    \item Atomic calculations of  $K$ and $F$ should report estimated uncertainties for these factors.  
    \item Whenever possible, the consistency between the theoretical and the King plot results should be assessed (combined approach). It should be noted that FS factors can often be obtained with reasonable accuracy using modern many-body methods~\cite{Bijaya1,Bijaya5,Bijaya6}.
    \item Isotope shift factors obtained from all available computational methods should be compared to evaluate their reliability. 
    \item It may be essential to consider higher-order IS factors in the analysis of nuclear charge radii derived from IS measurements, especially for higher $Z$.
    \item Isomeric charge radii should reference the respective ground state instead of, or in addition to, referencing them to a reference isotope.
\end{itemize}
%
To improve existing data -- or obtain more realistic uncertainties -- the following recommendations should be considered in future compilations  
\begin{itemize} 
    \item Assign a realistic uncertainty in those cases where the NMS has been estimated using the nonrelativistic scaling-law~\cite{Fricke:2004}.
    \item The accuracy of the Seltzer-coefficient approach for higher-order moments should be reconsidered and/or an alternative approach developed.
    \item Estimate uncertainties for the $V$ factors, whenever they are used.
\end{itemize}

\section{Multiply-charged ions}

\subsection{Light Helium-Like Ions}
\label{sec:heliumlike_ions}

The charge radius is, in principle, encoded in the frequency of an atomic transition, as long as the two states have significantly different probability densities inside the nuclear volume. Provided that a theoretical calculation of the transition frequency $\nu_\mathrm{theo}$, assuming a point-like nucleus, is available with an accuracy better than the field-shift contribution, the mean-square charge radius $\left\langle r_\mathrm{c}^2\right\rangle$ can be directly extracted according to 
\begin{equation}
    \left\langle r_\mathrm{c}^2\right\rangle = \frac{\nu_\mathrm{exp}-\nu_\mathrm{theo}}{F}.
\end{equation}
This does not require any additional information other than the calculated field shift factor $F$. While this is the standard approach for H-like systems and was applied to laser spectroscopy of muonic hydrogen~\cite{Pohl.2010}, deuterium~\cite{2016-Pohl}, and helium~\cite{Krauth.2021}, the extension to two-electron (He-like) systems is already extremely challenging, since it requires precision on the order of $10^{-9}$ or better in the transition frequency calculation for these complicated three-body systems. 
In a recent effort towards this goal, all QED contributions up to terms $m\alpha^7$ have been calculated for helium and helium-like systems~\cite{Yerokhin.2022}.  While the extraction of the nuclear charge radius became, in principle, possible for $1s2s\,^3\!S_1 \rightarrow 1s2p\,^3\!P_{0,1,2} $ transitions, there are still issues with the calculated transition frequencies, including $D$-states ~\cite{Patkos.2021}.

Precision spectroscopy has recently been applied to heavier helium-like atoms~\cite{Imgram.2023b}. A proof-of-principle experiment in $^{12}$C has resulted in a nuclear charge radius that was extracted from the transition frequency, which is in excellent agreement with the existing precise charge radii from elastic electron scattering as well as from muonic atom spectroscopy, but with a much larger uncertainty (See Fig.~\ref{fig:Abs}). This uncertainty is now dominated by the accuracy of the theoretical transition frequency; thus, any improvement on this side will reduce it.
There is obviously a need for QED calculations in helium-like ions, including dominant $m\alpha^8$ contributions. The experimental results from $^{12,13}$C$^{4+}$~\cite{Imgram.2023b,Muller.2025} can guide these calculations. Reaching the precision of experiment is a daunting task, but it could potentially lead to an accuracy comparable to the highest precision reached in muonic atom measurements with a crystal spectrometer~\cite{Ruckstuhl.1984}

Although absolute radius determinations in light systems with more than one electron are still challenging, the difference in mean-square charge radii $\delta \left \langle r^2 \right\rangle^{A, A'}$ between isotopes can be accurately extracted since all mass-independent terms cancel in the isotope shift, and the contributing mass shift can be calculated with 1--100\,kHz accuracy ($10^{-4}-10^{-6}$ relative precision for the mass shift) for up to five-electron systems~\cite{Maass.2019}. The most reliable approach to determine or improve the charge radii of light elements up to nitrogen in the near future is, therefore, a laser-spectroscopic measurement of the isotope shift in an appropriate charge state, in combination with the high accuracy determination of at least one stable isotope using muonic atoms, as followed in the quartet collaboration~\cite{2024-QUARTET}. Beyond nitrogen, the lifetime of the $1s2s\,^3\!S_1$ state in the helium-like system becomes too short ($<1$\,ms), and the transition frequencies of the transitions to the $1s2p\,^3\!P_{0,1,2}$ levels become shorter than 190\,nm.    

\subsection{Bound electron $g_j$-factors}
The gyromagnetic factor $g_j$ of electrons bound to a nucleus is sensitive to nuclear properties. The $g_j$-factor of hydrogen-like ions is a powerful probe of the nuclear charge radius as these systems are amenable to reliable theoretical calculations~\cite{2000_beier_g_j,2019-HLIS,PhysRevLett.120.043203,PhysRevLett.134.123001}. Recent assessments show that for differential $g_j$-factors, uncertainties in nuclear charge radii dominate the theoretical error budget across a broad range of nuclear charges ~\cite{2025-gfactorRadii}.

The determination of the $g_j$-factor is based on the precise measurement of the Larmor frequency of the bound electron, together with the measurement of the magnetic field via the cyclotron frequency of the ion~\cite{2019_Sturm}. Currently, the relative experimental precision of absolute $g_j$-factor measurements is at the low $10^{-11}$, primarily limited by magnetic field fluctuations~\cite{2013_Sturm, 2014_Sturm,2023_Morgner, 2023_Heisse}. Direct $g_j$-factor differences can be measured with two orders of magnitude more precision at low $10^{-13}$~\cite{2022_Sailer}.

Up to $Z \lesssim 40$, the QED binding corrections for the $g_j$-factor are calculated via a double series expansion in $\alpha$ and $Z \alpha$. The first order in $\alpha$ is calculated in all orders of $Z \alpha$ with high precision~\cite{2000_beier_g_j}. From $Z \geq 4$, the estimated uncertainty due to higher order QED effects in the second order of $\alpha$ dominates the theoretical prediction of the $g_j$-factor. This limits the absolute nuclear charge radius extraction since the effect of the finite nuclear size is small.
For example, in $g_j\left(^{28}\text{Si}^{13+}\right)$~\cite{2013_Sturm}, the finite nuclear size effect is at relative $1 \cdot 10^{-8}$, and the relative theory uncertainty is $3 \cdot 10^{-10}$~\cite{2017:PhysRevA.96.012502,2018_Czarnecki}, leading to a current charge-radius uncertainty of about $1.5\%$. This precision could be further improved by a factor of seven with improved theory. 

However, the $g_j$-factor difference for different isotopes of the same element can be calculated with much higher accuracy since the higher-order QED uncertainties are nearly identical and mostly cancel. This allows for extracting very precise differential radii. Recently, two hydrogen-like neon ions ($^{20}\text{Ne}^{9+}$ and $^{22}\text{Ne}^{9+}$) were co-trapped in the cryogenic Penning-trap setup of the \textsc{ALPHATRAP} experiment~\cite{2019_Sturm} and their bound electron isotopic $g_j$-factor difference has been measured 13 digits of precision relative to the absolute value~\cite{2022_Sailer}. These results improve the mean square nuclear charge radius difference of $^{20,22}\text{Ne}$ by a factor of nine compared to the literature value based on muonic atom spectroscopy~\cite{Angeli:2013epw}.

The extraction of the mean-square nuclear charge radius difference was experimentally limited by systematic uncertainties due to the magnetic field inhomogeneity and the anharmonicity of the electric trapping field. The main theoretical limitation is due to the effect of nuclear polarization of the two neon isotopes~\cite{1998_Mohr,2002_Nefiodov}. This contribution is evaluated using nuclear excitation models and constitutes an irreducible theoretical uncertainty, poses a hard limit on the precision of the theory to extract the mean square nuclear charge radius difference. Nevertheless, this theory uncertainty is currently an order of magnitude smaller compared to the experimental uncertainty~\cite{2025-gfactorRadii}.

For $Z > 40 $, the situation has very recently changed due to the completion of all order $Z \alpha$ calculations for the second order in $\alpha$~\cite{PhysRevLett.134.123001}. With the increased theoretical precision, it is possible to extract absolute nuclear charge radii of heavy ions.
As an example, we consider the $g_j$-factor of hydrogen-like tin $g_j \left(^{118}\text{Sn}^{49+}\right)$ which has been measured at the \textsc{ALPHATRAP} experiment~\cite{2023_Morgner}. 
$g_j\left(^{118}\text{Sn}^{49+}\right)$ is measured with a precision exceeding the current theoretical accuracy by about a factor of 40, with the experiment limited by temporal magnetic-field stability. Although a charge radius has not yet been extracted from the result, it is expected to be determined with an accuracy of $0.15\%$ (See Fig.~\ref{fig:Abs}). This is not much less accurate than a recent extraction from an advanced analysis of muonic atom energies in $^{120}$Sn~\cite{2025-Zr}.

In the future, it may be possible to extend the nuclear charge radius determination to nuclei with spin, enabling additional searches for physics beyond the Standard Model~\cite{2011_Yerokhin,2025_Quint}. Nuclear charge radius extractions based on bound electron $g_j$-factor measurements are virtually uncorrelated with any other type of nuclear charge radius measurement. Moreover, the uncertainties of the theory calculations for absolute and differential bound electron $g_j$-factors are themselves weakly correlated.
\subsection{Heavy Highly Charged Ions}
\label{sec:hci}
Extreme ultraviolet (EUV) and x-ray spectroscopy of highly charged ions (HCIs) offer an emerging new method for nuclear charge radius determinations, particularly in heavy elements and radioactive isotopes, where traditional methods such as muonic atom spectroscopy and elastic electron scattering face limitations. The technique relies on measuring the transition energies of Na-like and Mg-like ions, whose electronic configurations are simple enough to allow for highly accurate atomic structure calculations and can be conveniently produced in electron beam ion traps (EBITs)~\cite{Silwal2020, Hosier2024, Hosier2025, Staiger2025}. The current experimental and theoretical precision achievable for transition-energy separations is at the level of a few meV, corresponding to a nuclear charge-radius precision of roughly 0.01~fm, making the method competitive with traditional techniques for heavy nuclei (see Fig.~\ref{fig:diff}). 

By breeding ions from different elements into the same charge state, scaled differences in nuclear charge radii can be extracted between nuclides of different elements. These inter-element measurements allow for nuclear charge radius information to be shared across elements, reducing uncertainties and providing consistency tests of absolute radii from other techniques~\cite{2025-Endre, 2025-Staiger}. When applied to deformed nuclides, the spectra of highly charged ions do not exhibit the complex line splitting observed in muonic atom spectra, making HCIs particularly useful in deformed regions of the nuclear chart~\cite{2025-Endre}. 

The tension between theoretical results for the absolute value of Na-like $3p-3s$ transition energies has prevented the extraction of absolute charge radii directly from measured Na-like transition energies~\cite{Staiger2025}. Future work should seek to clarify the disagreement between relativistic many-body perturbation theory (RMBPT)~\cite{Gillaspy2013}, multiconfiguration Dirac-Hartree-Fock (MCDHF)~\cite{Staiger2025}, and $S$-matrix~\cite{2015_Sapirstein} calculations of $3p-3s$ transition energies. The current experimental precision of a few meV would lead to nuclear charge radius uncertainties near 0.01~fm if theoretical uncertainties were lowered and disagreements resolved.

Absolute radii have been determined from Li-like ions for $^{208}$Pb and $^{209}$Bi to precisions of 0.03~fm and 0.02~fm, respectively~\cite{2025_Yerokhin}. This work employed a technique similar to the combined muonic-electronic treatment described in Sec.~\ref{sec:combined_analysis} to account for higher-order nuclear moments. Re-measurement of the $2p-2s$ transition energies is required to lower these uncertainties further. 

\subsection{Recommendations}
To ensure that the conversion of HCI transition energies/energy differences to charge radius information is fully reproducible, we recommend that the following information be included in HCI radius publications:
\begin{itemize}

    \item Measured transition energies/energy differences, including correlations between them.

    \item Calculated transition energies/energy differences, including correlations between them.

    \item The nuclear charge density for which transition energies were calculated.
    
    \item Calculated nuclear sensitivity coefficients and their correlations, if relevant.
    
    \item The source of information on higher-order nuclear moments (e.g., electron scattering experiments, optical isotope shift studies, and nuclear density functional theory calculations).
    
\end{itemize}
Additionally, to ensure that absolute charge radii are reliable, we recommend additional theoretical efforts to calculate and compare Na-like and Li-like transition energies. The resolution of the Na-like tension, in particular, would immediately allow for the extraction of absolute radii.

\section{Outlook: Towards a new Compilation and Evaluation }
Beyond the analysis of specific measurement types, we recommend several general improvements in the compilation, evaluation, and dissemination of nuclear charge radii data. These efforts are intended to address the needs of producers, evaluators, and users of nuclear charge radius data.

An updated and comprehensive compilation of nuclear charge radii data is recommended. This compilation should follow the philosophy of \cite{Fricke:2004}, incorporating both experimental and theoretical information from original publications and unpublished theses. Wherever possible, primary sources should be consulted and verified directly rather than relying on intermediate reviews. The inclusion of reanalyses is also recommended, enabling the incorporation of revised uncertainties (e.g., from \cite{Fricke:2004, Angeli:1999, 2025-Mirror}) and updated theoretical inputs from atomic and nuclear structure calculations. The reanalysis should not replace the original data so that data traceability is maintained.

We recommend that compiled data be stored in a machine-readable format that enables flexible queries and supports the needs of evaluators. It is further recommended that it be accessible through a web-based interface, allowing users to identify available measurements, trace the inputs used in the extraction of specific charge radii, and determine whether the data have been reanalyzed or superseded. New publications should be included promptly, allowing interested users to stay up-to-date between major evaluations. The compiled data should not immediately impact recommended nuclear charge radii. Instead, any impact should come only after a critical evaluation at regular intervals.

A key development for future evaluations is the explicit inclusion of correlations among input data. Neglecting such correlations can lead to significant underestimation of uncertainties, particularly for elements with multiple stable isotopes. We therefore recommend that future evaluations employ a generalized least-squares approach, replacing the previously used weighted least-squares procedures. Correlations arising from shared theoretical inputs, experimental systematics, or common conversion factors should be propagated to the level of extracted radii whenever possible. When correlation information is unavailable, evaluators are encouraged to develop reasonable estimates in consultation with experts familiar with the specific techniques. The resulting covariance matrix of the recommended radii should be provided to enable proper uncertainty propagation by users.

Finally, the evaluation procedure itself should be transparent. We recommend that future evaluations clearly document which data sets have been included, what modifications or reanalyses have been performed, and how the recommended values have been obtained. Any assumptions introduced for completeness --- such as interpolation of $V_2$, SMS, or NMS values from neighboring systems --- should be explicitly stated, and the associated uncertainties quantified and incorporated. Radii derived from semi-empirical models or machine-learning approaches should be clearly identified as such, with distinct formatting to distinguish them from directly measured quantities.

\section*{Acknowledgment}
The authors thank the International Atomic Energy Agency for coordinating the meeting that facilitated the discussions leading to this publication.

\bibliography{apssamp}

\end{document}